\documentclass[12pt]{iopart}
\usepackage{latexsym}
\usepackage{fullpage}
\usepackage{graphics}
\usepackage{graphicx}

\begin{document}
\title{Plans for the LIGO-TAMA Joint Search for Gravitational Wave Bursts}

\author{
Patrick J.~Sutton$^{1}$,
Masaki Ando$^{2}$,
Patrick Brady$^{3}$,
Laura Cadonati$^{4}$,
Alessandra Di Credico$^{5}$,
Stephen Fairhurst$^{3}$,
Lee Samuel Finn$^{6}$,
Nobuyuki Kanda$^7$,
Erik Katsavounidis$^{4}$,
Sergey Klimenko$^{8}$,
Albert Lazzarini$^{1}$, 
Szabolcs Marka$^{1}$,
John W.~C.~McNabb$^{6}$,
Saikat Ray Majumder$^{3}$, 
Peter R.~Saulson$^{5}$,
Hideyuki Tagoshi$^9$,
Hirotaka Takahashi$^{9,10,11}$,
Ryutaro Takahashi$^{12}$,
Daisuke Tatsumi$^{12}$,
Yoshiki Tsunesada$^{12}$,
S.~E.~Whitcomb$^{1}$
}

\address{
${}^{1}$LIGO Laboratory, California Institute of Technology, 
Pasadena, CA 91125, USA}
\address{
${}^{2}$Department of Physics, University of Tokyo, Hongo, Bunkyo-ku, 
Tokyo 113-0033, Japan}
\address{
$^{3}$Department of Physics,
University of Wisconsin-Milwaukee, Milwaukee, WI 53201, USA}
\address{
$^{4}$LIGO Laboratory, Massachusetts Institute of Technology, 
Cambridge, MA 02139, USA}
\address{
$^{5}$Department of Physics, Syracuse University, Syracuse, NY 13244, USA}
\address{
$^{6}$Center for Gravitational Wave Physics, Department of
Physics, The Pennsylvania State University, University Park, PA
16802, USA }
\address{
${}^{7}$Department of Physics, Graduate School of Science, 
Osaka City University, Sumiyoshi-ku, Osaka, Osaka 558-8585, Japan}
\address{
${}^{8}$Department of Physics, University of Florida, Gainesville, 
FL 32611, USA}
\address{
${}^{9}$Department of Earth and Space Science, Graduate School of Science,
Osaka University, Toyonaka, Osaka 560-0043, Japan}
\address{
${}^{10}$Graduate School of Science and Technology,
Niigata University, Niigata, Niigata 950-2181, Japan}
\address{
${}^{11}$Yukawa Institute for Theoretical Physics, Kyoto University, Kyoto, 
Kyoto 606-8502, Japan}
\address{
${}^{12}$National Astronomical Observatory of Japan, Mitaka, Tokyo 181-8588, 
Japan \\
\texttt{e-mail:psutton@ligo.caltech.edu} 
}

\begin{abstract}
We describe the plans for a joint search for unmodelled 
gravitational wave bursts being carried out by the LIGO 
and TAMA collaborations using data collected during 
February-April 2003.  
We take a conservative approach to detection, 
requiring candidate gravitational wave bursts 
to be seen in coincidence 
by all four interferometers.
We focus on some of the complications of performing this 
coincidence analysis, in particular the effects of the 
different alignments and noise spectra of the interferometers.
\end{abstract}

\section{Introduction}
\label{sec:introduction}

At present several large-scale interferometric gravitational wave 
detectors are operating or are being commissioned: 
GEO \cite{Wi_etal:04}, 
LIGO \cite{Si:04}, 
TAMA \cite{Ta:04}, 
and Virgo \cite{Ac_etal:04}.
In addition, numerous resonant mass detectors have been operating 
for a number of years 
\cite{Al_etal:00,As_etal:03}.
Cooperative analyses by multiple observatories could be valuable
for making confident detections of gravitational waves and for 
extracting the most information from them.
This is particularly true for gravitational wave bursts (GWBs) 
from systems such as core-collapse supernovae 
\cite{ZwMu:97,DiFoMu:02a,DiFoMu:02b,OtBuLiWa:04},
black hole mergers \cite{FlHu:98a,FlHu:98b}, 
and gamma-ray bursters \cite{Me:02}, 
for which we have limited theoretical knowledge to guide us.
Advantages of coincident observations include a 
decreased rate of false alarms from random noise fluctuations, 
the ability to locate a source on the sky, 
the ability to extract polarization information, 
and better statistics on signal parameters \cite{Fi:01,Pa_etal:01,Ar_etal:03}.
Independent observations using different detector hardware and software 
also decrease the possibility of error or bias.

The TAMA and LIGO collaborations are currently conducting joint 
searches for gravitational waves in the LIGO Science Run 2 (S2) and 
TAMA Data Taking Run 8 (DT8) data, which was collected between February 14 
and April 14 2003 \cite{Si:04,Ta:04}.  
Three classes of gravitational wave transients are being sought:
\begin{enumerate}
\item
unmodelled GWBs from the gamma ray burst event GRB 030329 
\cite{Me:03,Pr_etal:03,Hj_etal:03,Mo_etal:04}; 
\item
inspiral signals from galactic binary neutron star systems (see 
for example \cite{S1inspiral,Ta_etal:04}); and
\item
unmodelled GWBs without an electromagnetic trigger (see for example 
\cite{S1burst,An:04}).
\end{enumerate}
In this article we report on the planned ``untriggered'' search 
for unmodelled GWBs.  This is the first 4-interferometer coincidence 
analysis for GWBs,\footnote{ 
See \cite{Al_etal:00,As_etal:03,Am_etal:89} for GWB searches involving 
networks of resonant mass detectors.}
involving detectors at three sites. 
We will focus on some of the challenges facing this joint search, 
in particular the differences in the noise spectra and in 
the alignments of the LIGO and TAMA detectors.  As we shall 
see, these motivate a coincidence search between LIGO and TAMA 
that involves a trade-off between the false alarm rate, the 
detection efficiency, and the bandwidth.
The result is a search for high-frequency GWBs (in the range 
[700-2000]Hz) characterized by a low false alarm rate.
We adopt this conservative approach to detection by the LIGO-TAMA 
detectors in order to gain experience sharing data and performing 
combined analyses.
This search could form a prototype for more 
comprehensive collaborative analyses in the future.

In Section~\ref{sec:procedure} we review the procedure used 
in the joint search for GWBs.  
In Section~\ref{sec:challenges} we examine some of the challenges facing the 
joint analysis and their implications.
In Section~\ref{sec:summary} we conclude with a few brief comments.

\section{Untriggered Bursts Analysis}
\label{sec:procedure}


The LIGO-TAMA joint search for GWBs is structured 
in the style of the LIGO S1 GWB analysis \cite{S1burst}
and the TAMA DT8 GWB analysis \cite{An:04}.  
Our goal is to detect GWBs or place upper limits on the 
rate of GWBs detectable by our instruments. 
Since we aim to be sensitive to the widest variety of 
waveforms which contain significant energy in the frequency 
range of our detectors, 
we choose to make minimal assumptions about the GWB waveforms; 
in particular we do not 
use templates or matched filters.
Instead, a variety of non-template based burst detection 
algorithms are employed to locate candidate GWBs 
\cite{An:04,Sy:02,AnBrCrFl:01,Kl_etal:04,Mc_etal:04}.

The analysis procedure is straightforward:
\begin{enumerate}
\item\label{etg}
Each collaboration independently generates lists of candidate GWBs.  
These events are characterized by (at minimum) 
a start and/or ``peak'' time, a duration, 
and an amplitude or SNR.  The LIGO events are also 
characterized by a frequency range, whereas the 
TAMA bursts detection scheme uses a fixed set of frequencies 
between 700Hz and 2000Hz \cite{An:04}. 
The event generation procedure may include 
vetoes of candidate events based on auxiliary data, such as 
from environmental monitors. 
In addition, 
events which are in coincidence in the three LIGO interferometers are 
subjected to a coherence test,  
in which the cross-correlation of the data from each pair of LIGO 
interferometers is required to exceed a certain threshold \cite{Ca:04}.
(Further study is required before including TAMA events in this 
coherence test due to the different alignment of TAMA; 
see Section~\ref{sec:antenna}.)
\item
The event lists from the four detectors 
are compared and candidate GWBs are selected 
by requiring coincident detection in each interferometer 
(LIGO and TAMA).
\item
The rate of accidental coincidences due to the background noise 
is estimated by repeating the coincidence and coherence tests 
after artificially 
shifting the time stamps of one or more lists of events.
It is intended that the event generation thresholds in step (\ref{etg}) 
will be kept high enough so that the average background rate will 
be less than one 
event over the observation time (approximately 250 hours). 
\item
An excess of coincident events above the background 
will be investigated as a possible detection; 
in the absence of an excess an  
upper limit on the rate of detectable events will be set 
using the background estimate.
\item
The efficiency of the LIGO-TAMA network at detecting real GWBs 
will be estimated by adding various simulated signals 
(astrophysical and ad-hoc) to the data in Monte Carlo experiments.
\end{enumerate}

It should be noted that the LIGO collaboration is also performing 
a separate coincidence search for GWBs using the LIGO S2 data alone below 
1.1kHz.  Candidate GWBs from the two analyses (if any) will be compared.


\section{Challenges for LIGO-TAMA}
\label{sec:challenges}

A coincidence analysis involving LIGO and TAMA is more complicated than, 
for example, a coincidence analysis involving the LIGO detectors only.  
The main complications are that the LIGO and TAMA detectors have 
different noise spectra and different alignments.



\subsection{Sensitivity vs.~Frequency}
\label{sec:spectra}

\begin{figure} 
  \begin{center}
  \includegraphics[width=12cm]{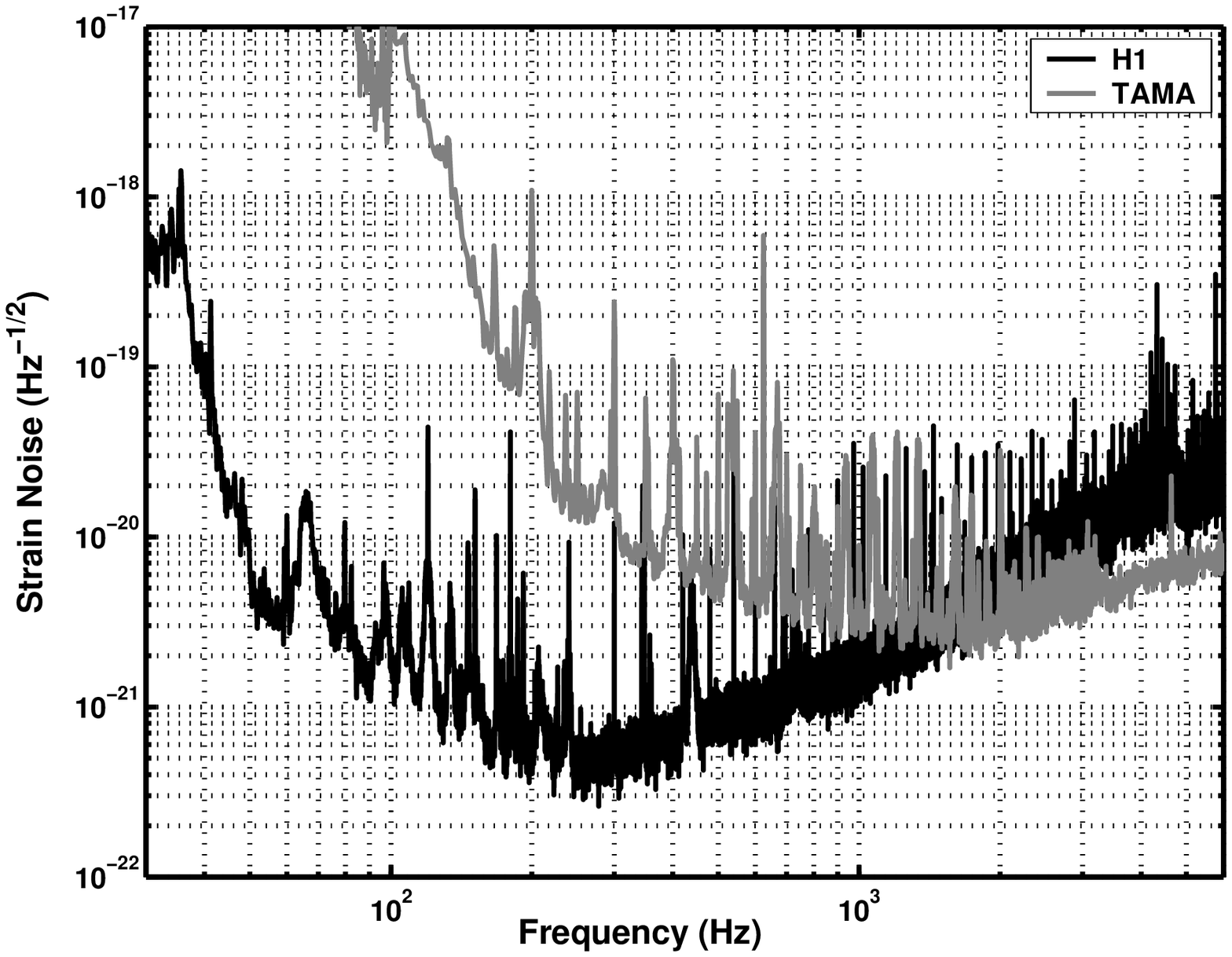}
  \caption{
\label{fig:spectra} 
Representative amplitude noise spectra for the LIGO-Hanford 4km (H1) and 
TAMA detectors.  Events from each detector are used in this analysis if  
they intersect the frequency range [700-2000]Hz, where each  
interferometer has approximately the same noise level.
}
  \end{center}
\end{figure}

Figure~\ref{fig:spectra} shows 
representative noise spectra from the LIGO-Hanford 
4km detector and the TAMA detector during S2/DT8 (the other LIGO detectors are 
similar to the one shown).  The individual detectors have maximum 
sensitivity at very different frequencies - about 250Hz for LIGO and 
1300Hz for TAMA.  

If we ignore differences in alignment, then requiring four-fold 
coincident detection 
means that the overall sensitivity of the network is limited 
by the least sensitive detector.  This is because 
our coincident analysis requires a signal to be detected 
by all interferometers; hence, we choose to focus our search on 
the minimum of the noise {\em envelope}, which occurs around 1000Hz.  
Specifically, we require that the frequency range of LIGO events 
overlap [700,2000]Hz; this is also the frequency range analyzed 
in the production of TAMA events.
Restricting the frequency range will reduce the rate of false 
alarms due to coincident noise fluctuations, while preserving
sensitivity to real GWBs that are detectable by 
both LIGO and TAMA.

Note that any low-frequency GWBs which are detectable by LIGO but not TAMA 
will be found in the LIGO collaboration's separate low-frequency search.
Thus, there is no danger of a detectable low-frequency GWB being 
missed due to our choice of frequency range for the LIGO-TAMA coincidence. 

\subsection{Antenna Patterns}
\label{sec:antenna}

\begin{figure} 
  \begin{center}
  \includegraphics[width=12cm]{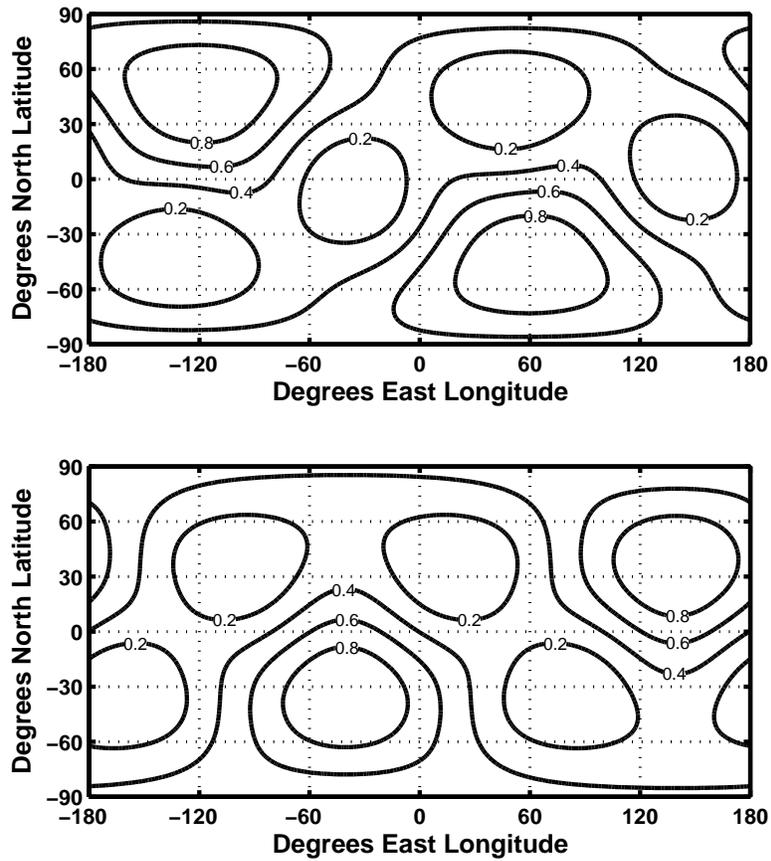}
  \caption{
\label{fig:antenna}
Antenna patterns $\rho$ (\ref{eq:rho}) in Earth-based coordinates 
(see for example \cite{Al_etal:01}).
The upper plot is for the LIGO Hanford detectors, while the lower plot 
is for TAMA.  
High contour values indicate sky directions of highest sensitivity.  The directions 
of LIGO's maximum sensitivity lie close to areas of TAMA's worst sensitivity 
and vice versa.
 }
  \end{center}
\end{figure}

Figure~\ref{fig:antenna} shows the antenna patterns 
\begin{equation}\label{eq:rho}
\rho \equiv F_+^2 + F_\times^2
\end{equation}
for the LIGO Hanford and TAMA detectors.  
(The LIGO Livingston antenna pattern is very similar 
to that of the LIGO Hanford detectors.)
These detectors are strongly non-aligned in that the areas of maximum 
sensitivity for LIGO overlap minima of TAMA sensitivity, 
and vice versa.  That is, LIGO and TAMA look with best 
sensitivity at different parts of the sky.  
%
This complicates LIGO-TAMA coincidence;
if we ignore differences in noise levels, then requiring four-fold 
coincident detection means that the overall sensitivity of the 
network to a given point on the sky 
is limited by the minimum of the antenna patterns.  

\begin{figure} 
  \begin{center}
  \includegraphics[width=12cm]{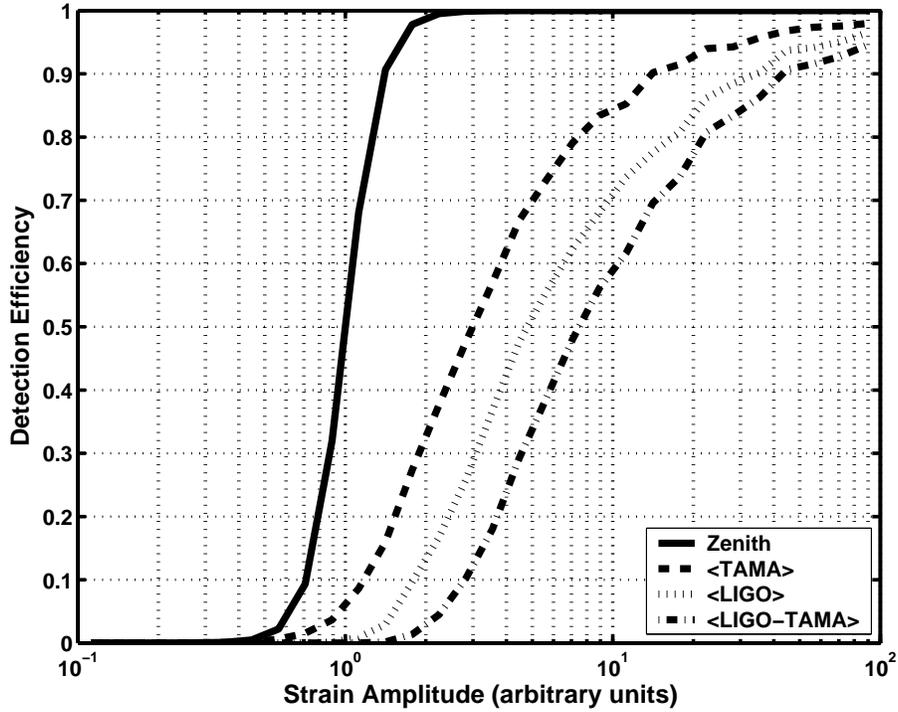}
  \caption{
\label{fig:eff} 
Sample efficiency estimates for the LIGO-TAMA network from Monte Carlo 
simulations.  Each interferometer is assumed to have identical 
sensitivity to a fiducial source (represented by the Zenith curve) 
when that source is optimally 
positioned and oriented with respect to the interferometer. 
The other curves show the efficiency to a population 
of such sources isotropically distributed over the sky and with 
random polarization.
The TAMA curve shows the detection efficiency of the TAMA detector 
alone, the LIGO curve shows the efficiency for coincident detection 
by all three 
LIGO detectors, and the LIGO-TAMA curve shows the efficiency 
for coincidence detection by TAMA and all three LIGO interferometers.
}
  \end{center}
\end{figure}

The impact of this alignment difference on the network sensitivity can 
be estimated via a simple Monte Carlo experiment.  
Let us assume each interferometer has the same sensitivity 
for some fiducial signal of interest.
We model this detection efficiency as a simple sigmoid in the logarithm 
of the signal amplitude, as shown in Figure~\ref{fig:eff}.\footnote{
We use equation (5.10) of \cite{S1burst} with $a=0.1$, $b=0$.
Choosing b=0 is the case in which all four detectors have the same sensitivity; 
this is approximately true around 1kHz (see Figure 1), and it is the simplest 
choice for illustrating the effect of the alignment separately from differences 
in the noise levels.
Sigmoid widths of order $a\simeq0.07$ to $a\simeq0.15$ are typical of the 
values found in the LIGO S1 analysis 
\cite{S1burst} and in the TAMA DT8 analysis.  
The network sensitivities depend only weakly on $a$.
}
We then generate linearly polarized signals of random amplitude 
and polarization angle, distributed isotropically across the sky.  
We project each signal onto the antenna pattern of each 
instrument, and compute the coincident detection 
efficiency for different combinations of 
interferometers.\footnote{See \cite{Ka:03} for the efficiency of 
an alternate network detection algorithm.}  
Figure~\ref{fig:eff} shows the resulting efficiencies as a function 
of the intrinsic signal amplitude.
For example, the minimum signal amplitude required for the three 
(nearly co-aligned) LIGO detectors to have a 50\% chance of 
simultaneous detection is approximately $4.7/2.9\simeq1.6$ times 
larger than the amplitude for 50\% detection by a single 
interferometer (exemplified here by TAMA).
The minimum amplitude for simultaneous detection by both 
LIGO and TAMA is approximately $2.6$ times the single-detector 
amplitude, or $1.6$ times the LIGO amplitude.  

The network efficiency is a function of the type of signal studied, 
and hence Figure 3 is only illustrative of what is expected for 
LIGO-TAMA.  Nevertheless, the loss in efficiency when adding detectors
to the network is generic when requiring coincident detection.  
Either collaboration could achieve better sky-averaged 
sensitivity by not requiring coincidence with the other.  
However, the loss in efficiency is accompanied by a drop in the 
false alarm rate.  For example, 
typical false-alarm rates at the single-interferometer level in 
this analysis are in the range $0.1-1$Hz.
Assuming a LIGO-TAMA coincidence window of order $0.1$s, 
each additional detector added to the network will lower the 
coincident false alarm rate by a factor of $10-100$.  
This false alarm reduction brings 
with it improved sensitivity to regions of the sky to which both LIGO 
and TAMA are simultaneously sensitive and is the principal 
scientific justification for our joint search.\footnote{
An alternate approach would be to lower the thresholds for 
event generation to keep the network false alarm rate constant 
as detectors are added to the network.  Whether the 
resulting gain in efficiency would be enough to offset 
the loss from adding differently aligned 
detectors depends on the relationship between the false alarm 
rate and efficiency for each detector.}

Another consequence of the difference in LIGO and TAMA alignments 
is that the detectors are sensitive to different polarization combinations 
of a gravitational wave.  This effect could be particularly important when
cross-correlating events between LIGO and TAMA.  Coherence tests have 
been shown to be a powerful means of reducing coincidence false alarm 
rates in LIGO \cite{Ca:04}, and extending these tests between LIGO 
and TAMA could reduce the network false alarm rate further.  This issue 
is under study.





\section{Summary}
\label{sec:summary}

The LIGO and TAMA collaborations are conducting joint analyses 
of the LIGO S2 - TAMA DT8 data collected during February-April 2003.  There 
are three analyses focusing on gravitational wave transients: 
unmodelled GWBs from GRB 030329, inspiral signals from galactic 
neutron star binaries, and an untriggered search for unmodelled GWBs. 

We have reviewed the plans for the untriggered search for GWBs and 
some of the complications of comparing LIGO and TAMA events. 
We take a conservative approach to detection, 
requiring candidate GWBs to be seen in coincidence 
by all four interferometers
at frequencies where all four  
have comparable sensitivity.
While this is expected to result 
in lower sensitivity of the network 
compared to the individual interferometers, it is the 
most conservative approach with respect to false alarms.

\section{Acknowledgments}

The LIGO Laboratory operates under Cooperative Agreement with the 
National Science Foundation, grant PHY-0107417.




\newpage

\newpage

\newpage

\end{document}